\documentstyle[12pt]{article}
\begin{document}

\baselineskip=24pt plus 2pt
\vskip -5cm
\hfill\hbox{NCKU-HEP/95-04}
\vskip -0.2cm
\hfill\hbox{27 September 1995}
\vskip 1cm

\begin{center}
{\large \bf Differential Renormalization of Scalar Field\\
Theory in the Background-Field Method}\\
\vspace{20mm}
Yaw-Hwang Chen, Min-Tsung He\\
and\\
Su-Long Nyeo{\footnote{Author to whom all correspondence should be addressed.\\
E-mail: t14269@MAIL.NCKU.EDU.TW.}}\\
\vspace{5mm}

Department of Physics, National Cheng Kung University \\
Tainan, Taiwan 701, R.O.C \\

\end{center}

\newpage

\begin{center}
{\bf ABSTRACT}
\end{center}
We introduce an approach for calculating the quantum loop corrections
in the $\phi^4$ theory.  Differential regularization and background-field
method are essential tools and are used to calculate the effective action
of the theory to two-loop order.  Our approach is considerably simpler
than other known methods and can be readily extended to higher-loop
calculations and to other models.
\newpage

Loop calculations in most quantum field theories are generally divergent
and a regularization procedure is thus needed.  The most commonly used method
is dimensional regularization, which respects gauge and Lorentz symmetries.
It is, however, not well suited to chiral theories because $\gamma_{5}$
is not well defined in arbitrary dimensions of space-time.

In this letter we consider an approach using differential regularization [1]
and background-field method [2] to calculate the effective action of the
$\phi^4$
theory to two-loop order.  The regularization method, which
was extensively studied in recent years [3], enjoys the fact that loop
calculations are carried out in coordinate space with well defined
$\gamma_{5}$.  Furthermore, the use of differential regularization
does not require counterterms.

We consider the $\phi^4$ theory in Euclidean space with action
\begin{equation}
S\left[ \phi \right] = \int d^4 x \left[-{1\over 2}\phi (x) \Box
\phi (x) + {\mu^2\over 2} \phi^2 (x) +
{\lambda\over 4!} \phi^4 (x) \right] \,\,,
\end{equation}
and generating functional
\begin{equation}
Z\left[ J \right] = \int D\phi \exp \left\{ - S[\phi] +
\int d^4 x J(x)\phi (x) \right\}\,\,.
\end{equation}
Let $\phi (x) = \phi_{cl} (x) + q(x)$ with $\phi_{cl} (x)$ being the
classical background field and $q(x)$ the quantum fluctuation field.
The classical field satisfies the equation of motion
\begin{equation}
-\Box \phi_{cl} (x) + \mu^2 \phi_{cl} (x) + {\lambda\over 3!}
\phi^{3}_{cl} (x) = J(x) \,\,.
\end{equation}
The generating functional now takes the form
\begin{equation}
Z\left[ J \right] = \int Dq \exp\left[ -S[\phi_{cl}]
-(S_{0} + S_{I} )
+ \int d^4 x J(x)\phi(x)\right] \,\,,
\end{equation}
where
\begin{eqnarray}
S_{0} & = & \int d^4 x \left[ -{1\over 2} q(x) \Box q(x)
  + {\mu^2\over 2} q^2 (x) + {\lambda\over 4} \phi^{2}_{cl}
  q^{2} (x) \right]\,\,,\\
S_{I} & = & \int d^4 x \left[ {\lambda\over 3!} \phi_{cl} (x) q^3 (x)
  + {\lambda\over 4!} q^4 (x) \right] \,\,.
\end{eqnarray}
To calculate the quantum loop corrections, we consider the effective action
\begin{equation}
\Gamma\left[{\overline\phi}\right] = \ln Z\left[ J\right] -
\int d^4 x J(x){\overline\phi}(x)\,\,,  {\overline\phi}(x)
\equiv {{\delta \ln Z}\over{\delta J(x)}}\,\,,
\end{equation}
and its loop expansion [2]
\begin{equation}
\Gamma\left[\phi_{cl}\right] = S^{(0)} + S^{(1)} + S^{(2)} + \cdots\,\,,
\end{equation}
where $S^{(0)} = S[\phi_{cl}]$, and $S^{(1)}$ and $S^{(2)}$ are the one-loop
and
two-loop effective actions, respectively,  which we shall calculate with our
approach.  For the calculations, we need the propagator $G(x,y)$ for the
quantum
field $q(x)$ in the background field $\phi_{cl} (x)$.  The propagator satisfies
\begin{equation}
\left(-\Box_{x} + \mu^2 + {\lambda\over 2} \phi^{2}_{cl} (x)\right)
G(x,y) = \delta^4 (x-y) \,\,.
\end{equation}
In our approach, we expand the propagator at short distance [4] as
\begin{equation}
G(x,y) = {1\over 4\pi^2} \left[ U(x,y){1\over z^2} +
         V(x,y) \ln z^2 M^2 + W(x,y) \right] \,\,,
\end{equation}
where $z^2 = (x-y)^2$ and $M$ plays the role of a subtraction point.
Substitution of (10) into (9) leads to
\begin{eqnarray}
& &U(x,y) \delta^4 (x-y) + {1\over 4\pi^2} \left\{ \ln z^2 \left( -\Box_{x}
V(x,y) + \left[ \mu^2 + {\lambda\over 2} \phi^{2}_{cl} (x) \right]
V(x,y) \right)\right.\nonumber\\
& & + {1\over z^2} \left( -\Box_{x} U(x,y)
+\left[ \mu^2 + {\lambda\over 2}\phi^{2}_{cl} (x) \right] U(x,y)
-4V(x,y) - 4z_{i} \partial_{i} V(x,y)\right) \nonumber\\
& &+{1\over z^4} 4z_{i} \partial_{i} U(x,y)
\left.+ \left( -\Box_{x} W(x,y) + \left[ \mu^2 + {\lambda\over 2}
\phi^{2}_{cl} (x) \right] W(x,y)\right)\right\}\nonumber\\
& &{\hskip 3cm}=\delta^4 (x-y)\,\,,
\end{eqnarray}
which can be rewritten as the following coupled equations:
\begin{eqnarray}
&&\delta^4 (x-y) U(x,y) = \delta^4 (x-y)\,,\\
&&-\Box_{x} U(x,y)
+\left[ \mu^2 + {\lambda\over 2}\phi^{2}_{cl} (x) \right] U(x,y)
-4V(x,y) - 4z_{i} \partial_{i} V(x,y) \nonumber\\
&&{\hskip 3cm}= z^2 g(x,y)\,,\\
&&4z_{i} \partial_{i} U(x,y) = z^4 f(x,y)\,,\\
&&-\Box_{x}
V(x,y) + \left[ \mu^2 + {\lambda\over 2} \phi^{2}_{cl} (x) \right]
V(x,y) = 0\,,\\
&&f(x,y) + g(x,y) + \left( -\Box_{x} W(x,y) + \left[ \mu^2
+ {\lambda\over 2} \phi^{2}_{cl} (x) \right] W(x,y)\right) = 0\,,{\hskip 1cm}
\end{eqnarray}
where $f(x,y)$ and $g(x,y)$ are unknown functions.  To solve for the
functions $\phi_{cl}(x)$, $f(x,y)$, $g(x,y)$
$U(x,y), V(x,y)$, and $W(x,y)$, we expand them in $x$ and $y$.
Solving (12)-(16), we obtain, in particular,
\begin{eqnarray}
U(x,y) & = & 1 + O(z^4 )\,\,,\\
V(x,y) & = & {1\over 4} (\mu^2 + {\lambda\over 2} \phi^{2}_{cl,0} )
             + {\lambda\over 8} \phi_{cl,0} \phi_{cl,1a_{1}}
             (x_{a_{1}} + y_{a_{1}}) + {3f^{0}_{0}\over 8}
             (2x\cdot y + y^2 )\nonumber\\
       &   & + {1\over 12} (\phi_{cl,0} \phi_{cl,a_{1} a_{2} }
             + 2\phi_{cl,1a_{1}} \phi_{cl,1a_{2}} )
             (x_{a_{1}} x_{a_{2}} + x_{a_{1}} y_{a_{2}} + y_{a_{1}}
             y_{a_{2}} )\nonumber\\
       &   & -{1\over 8} \left[ {1\over 6} ( \phi_{cl,0} \phi_{cl,2ii}
             +\phi_{cl,1i} \phi_{cl,1i})- {1\over 4} ( \mu^2
             +{\lambda\over 2} \phi^{2}_{cl,0})^2 \right] z^2\nonumber\\
       &   & +O(z^3 ) \,\,.
\end{eqnarray}
The propagator can now be written as
\begin{equation}
G(x,y)={1\over 4\pi^2} \left\{ {1\over z^2} + \left[ {1\over 4}
(\mu^2 + {\lambda\over 2} \phi^{2}_{cl,0} )
+{1\over 32} ( \mu^2 + {\lambda\over 2} \phi^{2}_{cl,0} )^2 z^2 \right]
\ln z^2 M^2 + \cdots \right\} \,.
\end{equation}

The one-loop effective action can be easily obtained from $G(x,y)$ [2]:
\begin{eqnarray}
S^{(1)} & = &{1\over 2}{\rm Tr} \ln G^{-1}\nonumber\\
        & = &{1\over 2}{\rm Tr} \ln G_{0}^{-1}
              +{1\over 2}\int d^4 x \delta \nu (x) \left[
               {\delta {\rm Tr}\ln G^{-1} \over \delta \nu (x) }
               \right]_{\nu =0}\nonumber\\
        &   &+{1\over 2}{1\over 2!} \int d^4 x d^4 y \delta \nu (x)
               \delta \nu (y) \left[
              {\delta^2 {\rm Tr}\ln G^{-1} \over \delta \nu (x)
               \delta \nu (y) } \right]_{\nu = 0}
              +\cdots\,\,,\nonumber\\
\end{eqnarray}
where
\begin{equation}
G^{-1} = \left[ -\Box_{x} + \mu^2 + {\lambda\over 2}
         \phi^{2}_{cl} (x) \right] \delta^4 (x-y)\,\,,
\end{equation}
\begin{equation}
{\rm Tr} \ln G^{-1}_{0} = {\rm Tr} \ln \left[ ( -\Box_{x} + \mu^2 )
\delta^4 (x-y) \right]\,\,,
\end{equation}
\begin{equation}
\nu (x) \equiv {\lambda\over 2} \phi^{2}_{cl} (x) \,\,.
\end{equation}
Hence, the one-loop effective action becomes
\begin{eqnarray}
S^{(1)} & = & {1\over 2} {\rm Tr} \ln G^{-1}_{0}\nonumber\\
        &   & +{1\over 2} \int d^4 x {\lambda\over 2}
               \phi^{2}_{cl} (x) {1\over 4\pi^2}
               \left[ {1\over z^2} + {\mu^2 \over 4} \ln z^2 M^2
               \right]_{z \approx 0}\nonumber\\
        &   & +{1\over 4} \int d^4 x ({\lambda\over 2}
                 \phi^{2}_{cl} (x) )^2 {1\over 4\pi^2}
                 \left[ {1\over 4} \ln z^2 M^2 \right]_{z \approx 0}
                 + \cdots \,\,,
\end{eqnarray}
which exhibits singularities at $z = 0$.  To remove these singularities,
we first consider the derivative of the action with respect to $\ln M^2$ [5]
\begin{equation}
{d S^{(1)}\over d \ln M^2}
= {\lambda\over 64\pi^2}\int d^4 x \mu^2 \phi^{2}_{cl}(x)
   +{\lambda^2\over 256\pi^2}\int d^4 x \phi^{4}_{cl}(x)\,\,.
\end{equation}
Integrating the expression with the introduction of a momentum scale
$\Lambda$, we obtain the effective action to one-loop order
\begin{eqnarray}
\Gamma\left[ \phi_{cl} \right] & \approx & S^{(0)}+S^{(1)} \nonumber\\
& = &\int d^4 x \left[ -{1\over 2}\phi_{cl}(x)\Box\phi_{cl}(x)
     +Z_{\mu}{\mu^2\over 2}\phi^{2}_{cl}(x)
     +Z_{\lambda} {\lambda\over
     4!}\phi^{4}_{cl}(x)\right]\,\,,\nonumber\\
\end{eqnarray}
where
\begin{eqnarray}
Z_{\mu} & = & 1 +{\lambda\over 32\pi^2} \ln{\Lambda^2\over M^2}\\
Z_{\lambda} & = & 1+{3\lambda\over 32\pi^2} \ln
             {\Lambda^2\over M^2}\,\,.
\end{eqnarray}
{}From the definition of the renormalized coupling constant
\begin{eqnarray}
\lambda_{R} & = & \lambda Z^{-1}_{\lambda} \nonumber\\
            & = & \lambda ( 1 - {3\lambda\over 32\pi^2}
                  \ln{\Lambda^2\over M^2})\,\,.
\end{eqnarray}
we have the expected one-loop $\beta$-function
\begin{eqnarray}
\beta (\lambda_{R})& = &M{\partial\over\partial M}\lambda_{R}\nonumber\\
           & = &  {3\lambda^{2}_{R} \over 16 \pi^2}\,.
\end{eqnarray}

We note that in our approach the calculation of the one-loop effective action
amounts to obtaining a short-distance expansion for the propagator in
the background field.  We did not employ differential regularization, but
a regularization is introduced in the expansion as an ansatz.

Next we shall calculate the two-loop effective action.
In the calculation we encounter two-loop divergences in the Feynman diagrams,
which must be handled by some regularization.
We shall use differential regularization, which automatically includes
the renormalization procedure and no counterterms are needed.
Its essential idea is to define the highly-singular terms in calculations by
\begin{equation}
{1\over (z^2 )^n } \ln^m (z^2 M^2 ) =
\underbrace{\Box\Box\cdots\Box}_{n-1}G(z^2)\,\,,
z^2 \neq 0\,\,, n \ge 2 \,, m \ge 0\,\,,
\end{equation}
and solve for the function $G(z^2)$, which has a well-defined Fourier
transform and depends on $2(n-1)$ integration constants.  Below we list
the regularized expressions that are used in the two-loop calculation:
\begin{eqnarray}
{1\over z^4} & = & -{1\over 4}
\Box {\ln(z^2 M^2 )\over z^2} \,\,, z^2 \neq 0\,,\\
{1\over z^6} & = & -{1\over 32}
\Box\Box {\ln( z^2 M^2)\over z^2} \,\,, z^2 \neq 0\,,\\
{1\over z^4}\ln(z^2 M^2) & = & -{1\over 8}
\Box { \left[ \ln(z^2 M^2)\right]^2 + 2 \ln( z^2 M^2) \over z^2}\,\,, z^2 \neq
0\,.
\end{eqnarray}
Here we have set the integration constants to be the subtraction point $M$,
which
amounts to adopting a renormalization scheme, and omitted the irrelevant
integration constants.

{}From Feynman rules, the two-loop effective action is given by
\begin{equation}
S^{(2)} = - 3\int d^4 x {-\lambda\over 4!} \left[ G(x,x) \right]^2
           - 3!\int d^4 x d^4 y ({-\lambda\over 3!})^2
                \phi_{cl} (x) \phi_{cl} (y) \left[ G(x,y)\right]^3\,\,.
\end{equation}
With the expansion for $G(x,y)$, we get
\begin{eqnarray}
S^{(2)} & = & - ({-3\lambda\over 4!16\pi^4})\int d^4 x \left[
              {1\over z^2} + {1\over 4} (\mu^2 + {\lambda\over 2}
               \phi^{2}_{cl})\ln z^2 M^2 +\cdots \right]^{2}_{z^2
               \approx 0}\nonumber\\
        &   & -{\lambda^2 \over 3! 128\pi^6 } \int d^4 x d^4 y
                \phi_{cl}(x) \phi_{cl}(y) \left[ {1\over z^6} +
               {3\over 4} (\mu^2 +{\lambda\over 2} \phi^{2}_{cl,0})
               {\ln z^2 M^2 \over z^4}\right.\nonumber\\
        &   & \left.+{{}\over{}} \cdots \right]_{z\approx 0}\,\,\nonumber\\
        & = & - ({-3\lambda\over 4!16\pi^4})\int d^4 x \left[
                    {1\over z^2} + {1\over 4} (\mu^2 + {\lambda\over 2}
                    \phi^{2}_{cl})\ln z^2 M^2 +\cdots \right]^{2}_{z^2
                    \approx 0}\nonumber\\
        &   & -{\lambda^2 \over 3! 128\pi^6 } \int d^4 x d^4 y
                    \phi_{cl}(x) \phi_{cl}(y) \left[-{1\over 32}
                    \Box\Box {\ln z^2 M^2 \over z^2}\right.\nonumber\\
        &   & \left. -{3\over 32}
             (\mu^2 + {\lambda\over 2} \phi^{2}_{cl,0})
              \Box{\ln^2 z^2 M^2 + 2\ln z^2 M^2 \over z^2}\right]_{z
              \approx 0}\,\,.
\end{eqnarray}
We next remove the singularities of the action at $z = 0$ by
considering the derivative of the action with respect to $\ln M^2$.
We have
\begin{eqnarray}
& &{d (S^{(1)} + S^{(2)} )\over d \ln M^2}\nonumber\\
& & = +{\lambda\over 64\pi^2}\int d^4 x \mu^2 \phi^{2}_{cl}
      +{\lambda^2\over 256\pi^2}\int d^4 x \phi^{4}_{cl}\nonumber\\
& &   +{1\over 16\pi^4}\int d^4 x \left[ {\lambda^2\over 192} {1\over 2}
      \phi_{cl} (x)\Box\phi_{cl} (x) + {\lambda^2\over 32} {\mu^2 \over 2}
      \phi^{2}_{cl}(x) +{3\lambda^3\over 16} {1\over 4!} \phi^{4}_{cl}(x)
      \right]\,\,.\nonumber\\
\end{eqnarray}
Integrating it with the introduction of a momentum scale $\Lambda$,
we obtain the renormalized effective action to two-loop order:
\begin{eqnarray}
\Gamma\left[ \phi_{cl} \right] & \approx & S^{(0)}+S^{(1)}+S^{(2)}\nonumber\\
& = &\int d^4 x \left[ -{Z_{\phi}\over 2}\phi_{cl}(x)\Box\phi_{cl}(x)
     +Z_{\mu}Z_{\phi}{\mu^2\over 2}\phi^{2}_{cl}(x)
     +Z_{\lambda} Z^{2}_{\phi}{\lambda\over
     4!}\phi^{4}_{cl}(x)\right]\,\,,\nonumber\\
\end{eqnarray}
where
\begin{eqnarray}
Z_{\phi} & = & 1+{1\over 16\pi^4}{\lambda^2\over 192}
             \ln{\Lambda^2\over M^2}\,\,,\\
Z_{\mu}Z_{\phi} & = & 1 +{\lambda\over 32\pi^2} \ln{\Lambda^2\over M^2}
             -{1\over 16\pi^4}{\lambda^2\over 32}
             \ln{\Lambda^2\over M^2}\,\,,\\
Z_{\lambda}Z^{2}_{\phi} & = & 1+{3\lambda\over 32\pi^2} \ln
             {\Lambda^2\over M^2}-{1\over 16\pi^4}{3\lambda^2\over 16}
             \ln{\Lambda^2\over M^2}\,\,.
\end{eqnarray}
Using the coupling constant renormalization
\begin{eqnarray}
\lambda_{R} & = & \lambda Z^{-1}_{\lambda} \nonumber\\
            & = & \lambda ( 1 - {3\lambda\over 32\pi^2}
                  \ln{\Lambda^2\over M^2}
                  +{1\over 16\pi^4}{17\lambda^2\over 96}
                  \ln{\Lambda^2\over M^2})\,\,,
\end{eqnarray}
we finally get the two-loop $\beta$-function and anomalous dimensions:
\begin{eqnarray}
\beta (\lambda_{R}) & = & M{\partial\over\partial M}\lambda_{R}\nonumber\\
            & = & {3\lambda^{2}_{R} \over 16 \pi^2} -
                  {17\lambda^{3}_{R} \over 768\pi^4}\,\,,\\
\gamma_{\mu}(\lambda_{R}) & = & {1\over Z_{\mu}}
       {\partial\over \partial \ln M^2} Z_{\mu}\nonumber\\
           & = &  {3\lambda_{R} \over 32 \pi^2} -
                  {7\lambda^{2}_{R} \over 3072\pi^4}\,\,,\\
\gamma (\lambda_{R}) & = & {\partial\over \partial \ln M^2} \ln Z_{\phi}
                             \nonumber\\
                     & = & {\lambda^{2}_{R}\over 3072\pi^4 }\,\,.
\end{eqnarray}
These are the identical results as that calculated more tediously
with dimensional regularization.

In summary, we have introduced an approach for calculating the quantum loop
corrections in the $\phi^4$ theory.  In this approach we considered a
short-distance expansion for the propagator in the background field.
The one-loop correction to the action can be easily derived from the
propagator.
To calculate the two-loop correction, we employed differential regularization,
which requires no counterterms; for renormalization is automatically a part
of the regularization procedure.  We easily obtained the well-known
two-loop $\beta$-function and anomalous dimensions.

Finally, we observe that for higher-loop order calculations, higher-order terms
in the short-distance expansions are needed.  Our approach can also be applied
to
gauge theories, but one should carefully take into account a gauge phase factor
for the difference of two space points in a short-distance expansion.

\vskip 1cm
\noindent{\large \bf Acknowledgments}

\noindent
This research was supported by the National Science Council of the Republic
of China under Contract Nos. NSC 84-2112-M-006-013 and NSC 85-2112-M-006-003.

\end{document}